\documentclass[12pt,a4paper]{article}

\newcommand{\blst}{1.24}
\renewcommand{\baselinestretch}{\blst}

\usepackage{amsmath}
\usepackage{amssymb}
\usepackage{a4wide}
\usepackage{epsfig}

\newcommand{\y}{\ensuremath{y}}
\newcommand{\z}{\ensuremath{z}}
\newcommand{\ig}{\ensuremath{ig}}

\newcommand{\piR}{\ensuremath{\pi R}}
\newcommand{\wt}[1]{\ensuremath{\widetilde{#1}}}
\newcommand{\dL}[1]{\ensuremath{\overleftarrow{\partial_{#1}}}}
\newcommand{\tr}{\ensuremath{\,\text{tr}}}
\newcommand{\omst}{\ensuremath{\overline{\omega}}}
\newcommand{\Mhat}{\ensuremath{\hat{M}}}
\newcommand{\mhat}{\ensuremath{\hat{m}}}
\newcommand{\vhat}{\ensuremath{\hat{v}}}
\newcommand{\N}{\ensuremath{\mathcal{N}}}
\newcommand{\Pd}{\ensuremath{\mathcal{P}}}
\newcommand{\F}{\ensuremath{\mathcal{F}}}
\newcommand{\J}{\ensuremath{\mathcal{J}}}
\newcommand{\Y}{\ensuremath{\mathcal{Y}}}

\begin{document}
\begin{titlepage}
  \begin{center}
    \hfill
    \begin{minipage}{6cm}
      \begin{flushright}
        TUM-HEP-560/04
      \end{flushright}
      \end{minipage}\\
    \bigskip
    \vspace{3\baselineskip}
  
    {\Large\bf Wilson Lines in Warped Space: Dynamical Symmetry Breaking 
and Restoration} \\
    \bigskip
    \bigskip
  
    {\bf  Kin-ya Oda} \\
    \smallskip
    {\small\sl  
      Physikalisches Institut der Universit\"at Bonn,
      D-53115 Bonn, Germany} \\
    \medskip
    {\bf Andreas Weiler} \\
    \smallskip
    {\small\sl
      Physik Department, Technische Universit\"at M\"unchen,
      D-85748 Garching, Germany} \\
    \bigskip
  
    {\tt odakin@th.physik.uni-bonn.de} \\
    {\tt Andreas.Weiler@ph.tum.de} \\
    \bigskip
    \vspace*{.5cm}
  
    {\bf Abstract} \\
    \end{center}
  \noindent
  The dynamics of Wilson lines integrated along a warped extra dimension has been unknown.
  We study a five dimensional $SU(N)$ pure gauge theory with Randall-Sundrum warped compactification on $S^1/Z_2$.
  We clarify the notion of large gauge transformations
  that are non-periodic on the covering space for this setup.
  We obtain Kaluza-Klein expansions of gauge and ghost fields
  for the most general twists and background gauge field configurations,
  which break the gauge symmetry at classical level in general.
  We calculate the one-loop effective potential and find that
  the symmetry corresponding to the subgroup allowing continuous Wilson lines is dynamically restored.
  The presented method can be directly applied to include extra fields.
  The connection to dynamical Scherk-Schwarz supersymmetry breaking in warped space is discussed.
  \bigskip
  \bigskip
  \end{titlepage}

\section{Introduction}
Warped compactification not only provides
a beautiful explanation how the large hierarchy
$m_\text{weak}/M_\text{Planck}$ is generated
from the exponential profile of its metric~\cite{Randall:1999ee}
but also is itself a quite general consequence of string theory
due to the fact that D-branes 
generically provide sources for warping
(See~\cite{WarpinString} and references therein).
In the original Randall-Sundrum model~\cite{Randall:1999ee},
five dimensional spacetime is compactified
on the orbifold $S^{1}/Z_{2}$, along which the
normalization of the four dimensional metric is exponentially scaled.
Orbifold compactification is a powerful tool in string theory to get
three generations and especially to reduce the rank of the gauge group
when combined with continuous Wilson lines~\cite{OrbwWil}. 
The values of these Wilson lines and the resulting gauge symmetry breaking pattern
must be determined dynamically
via the Hosotani mechanism~\cite{BunchofHosotani} 
when supersymmetry breaking is taken into account.
Studies of Wilson line dynamics on simpler orbifolds have been started
recently~\cite{Hosotanietal,Haba:2003ux}. 
These are also applied to gauge-Higgs unification models~\cite{GHunif}.

The orbifold compactification on $S^{1}/Z_{2}$
with radius $R=M_\text{GUT}^{-1}$ provides
a simple mechanism to solve the doublet-triplet splitting problem
in $SU(5)$ grand unified theories
(GUT's)~\cite{BunchofKawamura}. 
Realistic models along this line have been proposed
to implement low energy supersymmetry (SUSY) that is
broken at the weak scale~\cite{OrbGUT}.  
Once one considers an orbifold GUT,
it is tempting to think that SUSY is also broken by the compactification
via the Scherk-Schwarz mechanism~\cite{Scherk:1978ta}.
However, in such an orbifold model
the Scherk-Schwarz parameter, the amount of $SU(2)_{R}$ twist,
must be put to an extremely small value of order $m_\text{weak}/M_\text{GUT}$
at the classical level~\cite{Barbieri:2001yz}.
Furthermore, quantum corrections lead
either to restored supersymmetry $m_\text{SUSY}=0$
or to its violation of the order of $m_\text{SUSY}\simeq R^{-1}$,
unless one introduces an extra source of supersymmetry breaking
at the orbifold fixed point~\cite{vonGersdorff:2002tj}. 
This is generally true for symmetry breaking via the Hosotani mechanism in flat space.

Therefore it is natural to generalize the above considerations to 
a gauge theory in the bulk of the Randall-Sundrum geometry
where one can make use of the exponential hierarchy in its metric.
Warped SUSY GUT's are constructed in Refs.~\cite{Pomarol:2000hp,Goldberger:2002pc}
with SUSY breaking assumed to be by boundary conditions 
or by an extra source at the orbifold fixed point, respectively.  
Recently, it has been proposed
in the framework of superconformal gravity
that supersymmetry breaking by boundary conditions can be consistently defined and
equivalent to Wilson line breaking,
if the compensator multiplet has vanishing gauge coupling and
the warping is generated from the vacuum configuration of
the bulk scalars~\cite{Abe:2004ar}.\footnote{
    See also Ref.~\cite{WarpedSB} for discussions on
    the Scherk-Schwarz breaking in the usual warped setup.}
Eventually, the vacuum expectation value (vev) 
of the $SU(2)$ 
gauge field that is related to the Scherk-Schwarz twist
must be determined dynamically by quantum corrections to the effective potential.
To that end, it is
important to solve the dynamics of Wilson lines in
warped space, which has not been explored so far.\footnote{
    \renewcommand{\baselinestretch}{1}\footnotesize
    In Ref.~\cite{Contino:2003ve} a Wilson line in warped space is considered
    in the context of the AdS/Conformal Field Theory (CFT) correspondence,
    where the analysis is confined to a (potentially false) vacuum
    that corresponds to imposing both diagonal twists
    and vanishing background field configurations.
    See Ref.~\cite{KinyaAndi} for further discussions.}
\renewcommand{\baselinestretch}{\blst}\normalsize

In this paper we study the dynamics of Wilson lines of $SU(N)$ pure gauge theory
in the bulk of Randall-Sundrum geometry.
In the course of this calculation,
we derive for the first time
Kaluza-Klein expansions for the most general twists and background gauge field
configurations and calculate the corresponding one-loop effective potential.
In Section~\ref{flat_sec}, we briefly review the Hosotani mechanism on the orbifold $S^{1}/Z_{2}$.
In Section~\ref{KK_sec}, we obtain the Kaluza-Klein (KK) expansions of
the five dimensional gauge and ghost fields
with most general twists
in the presence of a gauge field background.
In Section~\ref{eff_pot_sec}, we calculate the effective potential for
the extra dimensional component of the background gauge field.
In the last section we summarize and discuss our result.

\section{\Large Wilson lines on flat $S^{1}/Z_{2}$}
\label{flat_sec}
We briefly review how twists and background gauge field configurations
are related by large gauge transformations.
We consider a five dimensional $SU(N)$ gauge theory
compactified on the orbifold $S^{1}/Z_{2}$,
which is obtained from the simply-connected space $R^{1}: -\infty<y<\infty$
by modding with $S^{1}$ and $Z_{2}$ identifications
  $y      \sim y+2\piR$ and
  $y      \sim -y$, where $R$ is the compactification radius.
Under these identifications, the gauge fields $A_{M}$ ($M=0,\dots,3;4$) are in general
twisted by global $SU(N)$ transformations
\begin{align}
  A_M(-y)      &= \pm P_0 A_M(y)      P_0^{-1}, \nonumber\\
  A_M(\piR+y)  &= \pm P_1 A_M(\piR-y) P_1^{-1}, &
  A_M(y+2\piR) &= UA_M(y)U^{-1},
  \end{align}
where
  the extra $\pm$~sign is positive for four dimensions and
    negative for the extra dimension.\footnote{
    In principle local identifications are possible
    but we assume them global for simplicity in this paper.
    }
(We use $\mu$ for $0,\dots,3$ and 
$y$ for both $x^{4}$ and index ``4'' such as $A_{\y}=A_{4}$.)
Note that the consistency conditions $U=P_1P_0$ and $P_0^2=P_1^2=1$ are imposed.
Starting from the most general twists 
we can always choose the following basis~\cite{Haba:2003ux}  
\begin{align}
  \setcounter{MaxMatrixCols}{20}
  \begin{array}{rclccrrrr}
      P_0 &=& \text{blockdiag}\,(\sigma_3,         &\ldots,&\sigma_3,         &I_r,&I_s, &-I_t,&-I_u),\\
      P_1 &=& \text{blockdiag}\,(\sigma_{\theta_1},&\ldots,&\sigma_{\theta_q},&I_r,&-I_s,&I_t, &-I_u),
    \end{array}
  \setcounter{MaxMatrixCols}{10}
  \label{bc_mat}
  \end{align}
where
  $I_{r}$ is $r\times r$ unit matrix,
  $\sigma_{a}$ ($a=1,2,3$) are Pauli matrices, and
  $\sigma_{\theta}
    =\sigma_{3}\cos{\theta}+\sigma_{1}\sin{\theta}
    =e^{-i\theta\sigma_{2}}\sigma_{3}
    =\sigma_{3}e^{i\theta\sigma_{2}}$.
($2q+r+s+t+u=N$.)
The $A_{\y}^{a}$ either within a block of $\pm I$
or connecting different blocks
does not have a zero-mode, a mode having vanishing KK mass,
and the dynamics of the corresponding Wilson line is trivial~\cite{Hall:2001tn}.\footnote{
    \renewcommand{\baselinestretch}{1}\footnotesize
    Of course when, say, $r<u$, we can combine $I_r$ and a part of $-I_u$
    to form $r$ additional $\sigma_\theta$ blocks with $\theta=0$.
    }
\renewcommand{\baselinestretch}{\blst}\normalsize
Therefore we can concentrate
on a $SU(2)$ subblock with twists
  $P_{0} = \sigma_{3}$ and
  $P_{1} = \sigma_{\theta}$ without loss of generality.
In general, only $A_{\y}^{(2)}$ ($A_{M}=A_{M}^{(a)}\sigma_{a}/2$)
has even $Z_{2}$ parities, hence a zero mode background:
  $gA^{(2)}_{\y}{}^{c} \equiv v$.
The KK expansions 
are given by
\begin{align}
  A^{(2)}_{\mu}(x,y) &=   \sum_{n=1}^{\infty}A^{(2)}_{\mu n}(x){\sin{ny\over R}\over\sqrt{\piR}}, &
  A^{(2)}_{\y}(x,y)  &= A^{(2)}_{\y0}(x){1\over\sqrt{2\piR}}
                         +\sum_{n=1}^{\infty}A^{(2)}_{\y n}(x) {\cos{ny\over R}\over\sqrt{\piR}}, \\
  A^{\pm}_{\mu}(x,y) &=      \sum_{n=-\infty}^{\infty}A_{\mu n}(x){e^{\pm i(m_{n}+v)y}\over N_{n}}, &
  A^{\pm}_{\y}(x,y)  &= \pm i\sum_{n=-\infty}^{\infty}A_{\y n}(x) {e^{\pm i(m_{n}+v)y}\over N_{n}},
                        \label{flat_Apm}
  \end{align}
where
  $A^{\pm}_{M} = {A^{(3)}_{M}\pm iA^{(1)}_{M}\over\sqrt{2}}$,  
  $A_{Mn}(x)$ are real fields,
  $m_{n}       = {n\over R}+{\theta\over\piR}-v$ are KK masses, and
  $N_{n}       = \sqrt{2\piR}\,m_{n}$ are normalization constants.
(Here we present a different form from literature in Eq.~\eqref{flat_Apm}
to make the orthogonality of wave functions transparent.)

Consider the following large gauge transformation that is non-periodic on the covering space $R^{1}$:
\begin{align}
  \ig A_{M}(y) &\rightarrow \ig\wt{A}_{M}(y) = \left[\Omega\left(\ig A_{M}-\dL{M}\right)\Omega^{-1}\right](y) &
  \text{with}\quad
  \Omega(y)    &=           \exp\left[i\varphi y{\sigma_{2}\over 2}\right],
  \label{large_gtf}
  \end{align}
which results in
  $g\wt{A}_{\y}^{(2)}{}^{c} = v-{\varphi/\piR}$.
We find that the shift of the background is canceled by the transformation of
$A^{\pm}_{M}\rightarrow \wt{A}^{\pm}_{M}=e^{\mp i\varphi y/\piR}A^{\pm}_{M}$
leaving its KK masses invariant: $\wt{m}_{n}=m_{n}$.
Now the new fields are twisted by matrices
  $\wt{P}_{0} = \sigma_{3}$ and
  $\wt{P}_{1} = \sigma_{\theta-\varphi} \equiv \sigma_{\tilde{\theta}}$.
Above, the two sets of twists $P_{i}$ and $\wt{P}_{i}$ with the corresponding backgrounds are
equivalent under the large gauge transformation,
by which one may e.g.\ diagonalize the twist $\tilde{\theta}=0$
or remove the background $\wt{A}_{\y}^{(2)}{}^{c}=0$, but not both.
If we choose to take the former (or latter) gauge,
different values of $\wt{A}_{\y}^{(2)}{}^{c}$ (or $\tilde{\theta}$) 
correspond to physically different vacua
which are degenerate at the classical level.
Quantum corrections determine whether the symmetry is dynamically
broken or restored, depending on 
the matter content~\cite{Hosotanietal,Haba:2003ux}.

\section{Kaluza-Klein expansions}
\label{KK_sec}
We consider a $SU(N)$ gauge theory in the bulk of
the Randall-Sundrum geometry~\cite{Randall:1999ee},
which is a five dimensional Anti de Sitter (AdS) space compactified on $S^{1}/Z_{2}$
with the metric:
\begin{align}
  G_{MN}dx^Mdx^N = e^{-2\sigma(y)}\eta_{\mu\nu}dx^\mu dx^\nu+dy^2,
  \end{align}
where
  $\eta_{\mu\nu}$ is the Lorentzian metric and
  $\sigma(y)$ is defined by $\sigma(y)=k|y|$ at $-\piR<y\leq\piR$,
  with $k$ being the inverse AdS curvature radius.
Elsewhere on the covering space $R^{1}$,
we define $\sigma$ by the periodicity condition $\sigma(y+2\piR)=\sigma(y)$.
For later use we also define\footnote{
    \renewcommand{\baselinestretch}{1}\footnotesize
    When there arises an ambiguity at the orbifold fixed point, say, around $y=0$,
    we can use the regularized form $\sigma(y)=k\delta\log\cosh(y/\delta)$
    with an infinitesimal $\delta=+0$ to check the expression.
    For our purpose we can use
      $\epsilon(y)     = \theta(y)-\theta(-y)$, 
      $\epsilon'(y)    = 2\left[\delta(y)-\delta(y-\piR)\right]$ and
      $\epsilon(y)^{2} = 1$
    at $-\piR<y\leq\piR$.
    \label{sig_ep}
    }
\renewcommand{\baselinestretch}{\blst}\normalsize
$\epsilon(y)=\sigma'(y)/k$ and $z(y)=e^{\sigma(y)}$.
We call the orbifold fixed points at $y=0$ and $y=\piR$ ultraviolet (UV)
and infrared (IR) branes, respectively, and write $\z_{0}=\z(0)=1$ and $\z_{1}=\z(\piR)=e^{\pi kR}$.
In this paper we assume that the radion is already stabilized e.g.\  by
the Goldberger-Wise mechanism~\cite{Goldberger:1999uk}.

We employ the background field method,
separating the gauge field into classical and quantum parts
$A_{M} = A_{M}^{c}+A_{M}'$,
and take the following gauge fixing\footnote{
    The choice $t=4$ and $\xi=1$ makes
    $f$ manifestly invariant under five dimensional diffeomorphisms.
    }
\begin{align}
  S_{f} &= -{1\over\xi}\int d^{4}x\int_{-\piR}^{\piR}dy\sqrt{-G}\tr\left[ff\right], &
  \text{with}\quad
  f     &= \z^{2}\eta^{\mu\nu}D_{\mu}^{c}A_{\nu}'
           +\xi \z^{t}D_{\y}^{c}\z^{-t}A_{\y}',
  \end{align}
where
  $D_{M}$ is the gauge covariant derivative and
  the superscript $c$ indicates that the gauge field is replaced by its classical part,
  e.g.\ $D_{M}^{c}A_{N}'=\partial_{M}A_{N}'+\ig[A_{M}^{c},A_{N}']$.
We consider the pure gauge background $F_{MN}^{c}=0$,
being a classical potential minimum,
and assume $A_{\mu}^{c}=0$
since it can be gauged away towards spatial infinity.
When we choose $\xi=1$ and $t=2$, the quadratic terms for
the gauge and ghost fields simplify: 
\begin{align}\label{quadraticaction}
  S &= \int d^{4}x\int_{-\piR}^{\piR}dy\tr\left[
          \eta^{\mu\nu}A_\mu'(\Box+\Pd_{4})A_\nu'
         +A_\y'\z^{-2}(\Box+\Pd_{\y})A_\y'
         +2\z^{-2}\omst'(\Box+\Pd_{4})\omega'
         \right],
  \end{align}
where
  $\Pd_{4}=D_\y^c\z^{-2}D_\y^c$,
  $\Pd_{\y}=D_\y^cD_{\y}^{c}\z^{-2}$, and
  we have put $\omega^{c}=\omst^{c}=0$.
(The surface terms vanish consistently for the given boundary conditions below.)

Following the same argument as in the flat case,
we concentrate on a $SU(2)$ subblock with twists
  $P_{0} = \sigma_{3}$ and
  $P_{1} = \sigma_{\theta}$, without loss of generality.
Again a zero mode resides only in $A_{\y}^{(2)}$ and we can write
  $gA_{\y}^{(2)}{}^{c}(y) = v\z^{2}$.
(The derivation of the form of zero mode $\z^{2}$ is given below.)
To obtain the KK expansions, we follow the strategy of Ref.~\cite{Gherghetta:2000qt}.
First, we solve the bulk KK equations at $0<y<\piR$ in terms of $\z$ neglecting all the boundary effects.
Second, we put the boundary conditions at $\z=\z_{0}$ and $\z_{1}$
on the obtained ``downstairs'' solution
to make it consistent with the $Z_{2}$ twists
so that the ``upstairs'' field on the covering space
is well-defined, i.e. continuous everywhere.

Let us start with $A_{\mu}^{(2)}$ and $A_{\y}^{(2)}$ which have definite odd and even $Z_{2}$ parities.
We obtain the following KK expansions
\begin{align}
  A_\mu^{(2)}(x,y)
    &= \sum_{n=1}^\infty A_{\mu n}^{(2)}(x)\epsilon
{\z[J_1(\Mhat_n\z)+B_{n}Y_1(\Mhat_n\z)]\over N_n}, &
  A_\y^{(2)}(x,y)
    &= \sum_{n=0}^\infty A_{\y n}^{(2)}(x)
 {f_{n}(\z)\over \N_n},
  \end{align}
where
  $N_n$, $\N_{n}$ are normalization constants,
  $B_{n} = -{J_1(\Mhat_n\z_0)\over Y_1(\Mhat_n\z_0)}= -{J_1(\Mhat_n\z_1)\over Y_1(\Mhat_n\z_1)}$, and
  the downstairs KK wave functions for vectoscalar
    $f_{n}(\z)=\z^{2}[J_{0}(\Mhat_{n}\z)+B_{n}Y_{0}(\Mhat_{n}\z)]$
    are defined for later use.
The KK masses $M_{n}=k\Mhat_{n}$ are determined by zeros of the KK mass function:
$J_{1}(\Mhat\z_{1})Y_{1}(\Mhat\z_{0})-Y_{1}(\Mhat\z_{1})J_{1}(\Mhat\z_{0})$,
which we find is exactly the same for both $A_{\mu}^{(2)}$ and $A_{\y}^{(2)}$.

In order to get the KK expansions of $A_{M}^{\pm}$, it is convenient to perform
the large (background) gauge transformation~\eqref{large_gtf},
which again results in the new twists
  $\wt{P}_{0} = \sigma_{3}$,
  $\wt{P}_{1} = \sigma_{\theta-\varphi} \equiv \sigma_{\tilde{\theta}}$
and the background
\begin{align}
  g\wt{A}_{\y}^{(2)}{}^{c}
    &= v\z^{2}-{\varphi\over\piR}
     = \left(v-{2ka^{2}\varphi\over 1- a^{2}}\right)\z^{2}
       -\frac{\varphi}{\pi R}\sum_{n=1}^{\infty}{(1,f_{n})\over\N_{n}^{2}}f_{n}(\z),
       \label{with_higher_modes}
  \end{align}
where
  $a=\z_{0}/\z_{1}=e^{-k\piR}\ll 1$ and
  $(1,f_{n})=\int_{\z_{0}}^{\z_{1}}2{d\z\over k\z}\z^{-2}f_{n}(\z)$.
Unlike in the flat case, all the higher KK modes are generated, as can be seen in the last step of
Eq.~\eqref{with_higher_modes}.
However, these higher modes all vanish
when integrated along the extra dimension in the Wilson line 
due to the orthogonality conditions of KK wave functions
and one would expect that these modes can be gauged away.

To see this, consider the following background gauge transformation
  $\check\Omega(y) = \exp\left[i\F(y){\sigma_{2}\over 2}\right]$,
which we require to be normal in the sense that $\F(y)$ is continuous everywhere and periodic $\F(y+2\piR)=\F(y)$.
When $\F(y)$ is odd $\F(-y)=-\F(y)$, twists $P_{i}$ are left
invariant under this transformation, while
gauge fields transform as
  $gA_{\y}^{(2)} \rightarrow g\wt{A}^{(2)}_\y  = gA^{(2)}_\y-\F'(y)$.
Again, this shift is canceled by the transformation of $A^{\pm}_{M}$ in its KK mass.
Let us take
\begin{align}
  \F(y)    &= k\sum_{n=1}^{\infty}\varphi_{n}F_{n}(y), &
  \text{with}\quad
  F_{n}(y) &= \int_{0}^{y}dy' f_{n}\left(z(y')\right),
  \end{align}
where the summation is over all the non-zero modes.
By definition, $F_{n}(y)$ is odd and its derivative is $f_{n}(y)$.
Due to the downstairs boundary conditions, we find that $F_{n}(y)$ vanishes at both boundaries,
i.e., the transformation $\check\Omega(y)$ is continuous everywhere on the covering space
as well as periodic, as promised.
To summarize, all the non-zero mode can always be removed
by taking $\varphi_{n}$ appropriately,
without changing the twists $P_{i}$.

Now we choose $\varphi=\theta$ to diagonalize the twist $\tilde{\theta}=0$
and gauge away all the resulting non-zero mode background.
We then have definite $Z_{2}$ parity for all $\wt{A}_{M}^{(a)}$.
Hereafter we omit the tilde for notational simplicity.
The KK expansions are obtained as\footnote{
    To derive Eq.~\eqref{KK_Apm}
    we have used $\epsilon^{2}=1$ in $A_{\y}^{(1)}$,
    which can be justified similarly as above.
    }
\begin{align}
  A^{\pm}_{\mu}(x,y)
    &= \sum_{n=0}^{\infty}A_{\mu n}(x)
       E^{\pm i\epsilon{\vhat\z^{2}/ 2}}{\chi_{1,n}^{\pm}(\z)\over N_{n}}, &
  A^{\pm}_{\y}(x,y)
    &= \sum_{n=0}^{\infty}A_{\y n}(x)
       \epsilon E^{\pm i\epsilon{\vhat\z^{2}/ 2}}{\chi_{0,n}^{\pm}(\z)\over\N_{n}},
       \label{KK_Apm}
  \end{align}
where
  $A_{Mn}(x)$ are real fields,
  $\vhat={v/k}-{2\theta a^{2}/(1-a^{2})}$ is the dimensionless vev,
  $E^{\pm i\epsilon w}\equiv\cos w \pm i\epsilon\sin w$,
  $N_{n}$ and $\N_{n}$ are normalization constants, and
  $\chi_{\nu,n}^{\pm}(\z)=\z^{2-\nu}[\alpha^{\pm}_{n}J_{\nu}(\mhat_{n}\z)+\beta^{\pm}_{n}Y_{\nu}(\mhat_{n}\z)]$
  are the downstairs KK wave functions
  with $\mhat_n \equiv m_n/k$ and $\alpha_n^\pm,\beta_n^\pm$ being dimensionless KK masses and complex constants, respectively.
  (Note $\alpha^{\pm}_{n}={\alpha^{(3)}_{n}\pm i\alpha^{(1)}_{n}\over\sqrt{2}}$ etc.)
The boundary conditions on the downstairs fields can be summarized as
\begin{align}
  M(\mhat_{n})\vec{V}=0,
  \end{align}
with
$\vec{V}=(\alpha^{(3)}_{n}\,\beta^{(3)}_{n}\,\alpha^{(1)}_{n}\,\beta^{(1)}_{n})^{T}$
for $A_{\mu}$,
$\vec{V}=(\alpha^{(1)}_{n}\,\beta^{(1)}_{n}\,\alpha^{(3)}_{n}\,\beta^{(3)}_{n})^{T}$
for $A_{\y}$, and
\begin{align*}
  \lefteqn{M(\mhat_{n})=}\nonumber\\
  &
      \begin{pmatrix}
        \J_{C}(1)        & \Y_{C}(1)        & \mp \J_{S}(1) & \mp \Y_{S}(1) \\
        \J_{C}(0)        & \Y_{C}(0)        & \mp \J_{S}(0) & \mp \Y_{S}(0) \\
        \pm\sin\frac{\vhat\z_{1}^{2}}{2} \, J_{\nu}(\mhat_{n}\z_{1})  & \pm\sin\frac{\vhat\z_{1}^{2}}{2}\, Y_{\nu}(\mhat_{n}\z_{1})  &
           \cos\frac{\vhat\z_{1}^{2}}{2}\, J_{\nu}(\mhat_{n}\z_{1})  &    \cos\frac{\vhat\z_{1}^{2}}{2}\, Y_{\nu}(\mhat_{n}\z_{1}) 
        \\
        \pm\sin\frac{\vhat\z_{0}^{2}}{2}\, J_{\nu}(\mhat_{n}\z_{0})  & \pm\sin\frac{\vhat\z_{0}^{2}}{2}\, Y_{\nu}(\mhat_{n}\z_{0})  &
           \cos\frac{\vhat\z_{0}^{2}}{2}\, J_{\nu}(\mhat_{n}\z_{0})  &    \cos\frac{\vhat\z_{0}^{2}}{2}\, Y_{\nu}(\mhat_{n}\z_{0}) 
        \end{pmatrix},
  \end{align*}
where
  upper and lower signs as well as $\nu=1$ and 0 are for $A_{\mu}$ and $A_{\y}$, respectively,
    $\begin{pmatrix}\J_{C}(i)\\ \J_{S}(i)\end{pmatrix}
        = \begin{pmatrix} \cos\frac{\vhat\z_{i}^{2}}{2} & -\sin\frac{\vhat\z_{i}^{2}}{2}\\
                          \sin\frac{\vhat\z_{i}^{2}}{2} &  \cos\frac{\vhat\z_{i}^{2}}{2}\end{pmatrix}
          \begin{pmatrix} \nu J_{\nu}(\mhat\z_{i})+\mhat_{n}\z_{i}J'_{\nu}(\mhat\z_{i}) \\
                          \vhat\z_{i}^{2} J_{\nu}(\mhat\z_{i}) \end{pmatrix}$,
  and $\Y_{C}$ and $\Y_{S}$ are defined similarly to $\J_{C}$ and $\J_{S}$ with $J$ replaced by $Y$.
  ($J_{\nu}$ and $Y_{\nu}$ are Bessel functions of order $\nu$.)
The KK mass function for $A^{\pm}$ is obtained from the determinant of the boundary condition matrix:
\begin{align}\label{massfunction}
  N(\mhat)  &= \det M(\mhat),
  \end{align}
and $N(\mhat) = 0$ determines the KK masses.
We find that the $N$'s are exactly the same for $A_{\mu}$ and $A_{\y}$
and that its dependence on $\vhat$ is only through the term $\frac{2}{\pi^2}\cos[\vhat(\z_{1}^{2}-\z_{0}^{2})]$.
The KK expansions for the ghost field is obtained similarly to
$A_{\mu}$.  (Recall that the antighost field $\omst$ is not
necessarily the complex conjugate of the ghost field $\omega$.)

\begin{figure}
  \begin{center}
    \leavevmode
    \epsfig{file=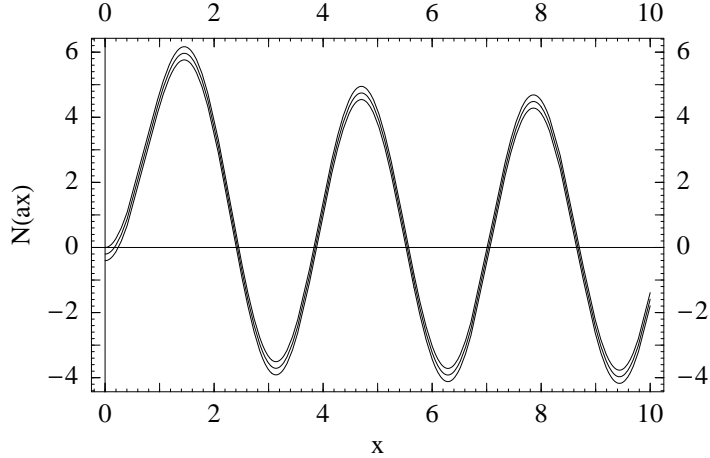,width=10cm}
    \caption{KK mass function $N(ax)$ vs $x$
             with $a=10^{-15}$ for ${\vhat/ a^{2}}=0,{\pi\over 2},\pi$.
             The zeros of $N$ correspond to the KK masses.
             Note that a massless mode appears only for $\vhat=0$ mod $2\pi$.}
    \label{fig_nfunc}
    \end{center}
  \end{figure}

In Fig.~\ref{fig_nfunc} we plot $N(ax)$ as a function of $x=\mhat/a$
within half a period of $\vhat$, i.e.\ for $\vhat(\z_{1}^{2}-\z_{0}^{2})=0,\pi/2,\pi$ from above to below
when $a=10^{-15}$.
The points where the curve crosses the $x$-axis give the values of the corresponding KK masses,
with even and odd modes appearing in alternating order.
We can see that there appears an extra massless mode for $\vhat=0$, as expected, and
that the dependence on the variation of $\vhat$ is strongest for this
would-be zero mode,
whose mass can
be easily determined from Eq.~\eqref{massfunction} to be
\begin{align}
  m_0 &= ka\sqrt{1-\cos(\vhat/a^2) \over k \pi R},
  \end{align}
in the approximation $a\ll 1$. For the maximal breaking with $\cos(\vhat/a^2) = -1$,
    we find $m_0\approx 0.24\,ka$ for $a=10^{-15}$.

\section{One loop effective potential}
\label{eff_pot_sec}
Following convention,
we perform the dimensional reduction
in the coordinate frame where the warp factor is unity at the UV brane.\footnote{
    The physical effective potential from the point of view of the IR brane
    will be enhanced by $a^{-4}$~\cite{Garriga:2000jb}.
    }
The contribution of a pair of $Z_{2}$ even and odd gauge fields $A^{(3)}_{M}$ and $A^{(1)}_{M}$ is
\begin{align}
  V_\text{eff}
    &= {\mu^{4-d}\over 2}\int{d^dp\over(2\pi)^d}\sum_{n=0}^{\infty}\log(p^2+m_n^2)
     = -{1\over 2}{(ka)^4\over(4\pi)^2}\left(4\pi\mu^2\over k^2a^2\right)^{\varepsilon/2}
       \Gamma\left(-2+{\varepsilon\over 2}\right)\sum_{n=0}^{\infty} x_n^{4-\varepsilon},
  \end{align}
per degree of freedom, where
  $d=4-\varepsilon$ is the number of dimensions with $\varepsilon$ being infinitesimal,
  $\mu$ is an arbitrary scale, and
  $x_n=m_n/ka$ is the dimensionless KK mass.
The infinite sum over KK masses can be evaluated utilizing
zeta function regularization techniques~\cite{Garriga:2000jb,Toms:2000bh,Goldberger:2000dv}
\begin{align}
  v_\text{eff}(\vhat)
    &\equiv -\Gamma\left(-2+{\varepsilon\over 2}\right)\sum_{n=0}^{\infty} x_n^{4-\varepsilon}
     =      -\Gamma\left(-2+{\varepsilon\over 2}\right)
             \int_C{dx\over 2\pi i}x^{4-\varepsilon}{N'(ax)\over N(ax)},
  \end{align}
where $C$ is a contour encircling all the poles on the positive real
axis counter-clockwise.
Note that these are the only poles in the right half plane
since there is a one-to-one correspondence between the zeros of the
KK mass function~\eqref{massfunction} and the eigenvalues of
the operators $\mathcal{P}_4$ and $\mathcal{P}_\y$ in Eq.~\eqref{quadraticaction}
which are Hermitian with respect to our boundary conditions.

After a few manipulations, we find
\begin{align}
  v_\text{eff}(\vhat)
    &= I_\text{IR}+{I_\text{UV}\over a^{4-\varepsilon}}
       +2\int_{0}^{\infty}dx\,x^{3-\varepsilon}\log\Bigg[\nonumber\\
    &\quad\mbox{}
         1-{1\over 2}\left({K_{0}(x)I_{0}(ax)\over I_{0}(x)K_{0}(ax)}
                          +{K_{1}(x)I_{1}(ax)\over I_{1}(x)K_{1}(ax)}
                          -{K_{0}(x)I_{1}(ax)\over I_{0}(x)K_{1}(ax)}
                          -{K_{1}(x)I_{0}(ax)\over I_{1}(x)K_{0}(ax)}\right)\nonumber\\
    &\quad\mbox{}
         +{K_{0}(x)K_{1}(x)I_{0}(ax)I_{1}(ax)\over I_{0}(x)I_{1}(x)K_{0}(ax)K_{1}(ax)}
         -{\cos\left({\vhat\over a^{2}}(1-a^{2})\right)\over 2ax^{2}I_{0}(x)I_{1}(x)K_{0}(ax)K_{1}(ax)}\Bigg],
         \nonumber\\
    &\simeq I_\text{IR}+{I_\text{UV}\over a^{4-\varepsilon}}
       +2\int_{0}^{\infty}dx\,x^{3-\varepsilon}\log\bigg[
         1-{I_{0}(x)K_{1}(x)-K_{0}(x)I_{1}(x)-{1\over x}\cos{\vhat\over a^{2}}  \over
           2I_{0}(x)I_{1}(x)\left(\gamma+\log{ax\over 2}\right)}\bigg],
  \label{eq_Veff_vec}
  \end{align}
where divergent integrals $I_\text{IR}$ and $I_\text{UV}$ are independent of $v$ and $a$
and can be absorbed in the renormalization of the IR- and UV-brane tensions, respectively,
as in Ref.~\cite{Goldberger:2000dv}.
($I_{\nu}$ and $K_{\nu}$ are the modified Bessel functions.)
We find that the effective potential is a periodic function of $\vhat$ with the period $2\pi a^{2}/(1-a^{2})$
as is expected from the shape of the KK mass function.
In the last line of Eq.~\eqref{eq_Veff_vec}, the small $a$ limit is taken,
assuming that
$\vhat$ is within the first period, i.e.\ $\vhat/a^{2}=O(1)$,
without loss of generality.\footnote{
    \renewcommand{\baselinestretch}{1}\footnotesize
    For large $x$, the integrand goes to zero and a small $a$ expansion can be performed
    with converging coefficients.}
\renewcommand{\baselinestretch}{\blst}\normalsize
(One might find it suggestive that the scale of the period of $v$ is of the order of $ka^{2}$,
which roughly corresponds to the order of the observed value of the cosmological constant $\simeq \text{meV}$.)

\begin{figure}
  \begin{center}
    \leavevmode
    \epsfig{file=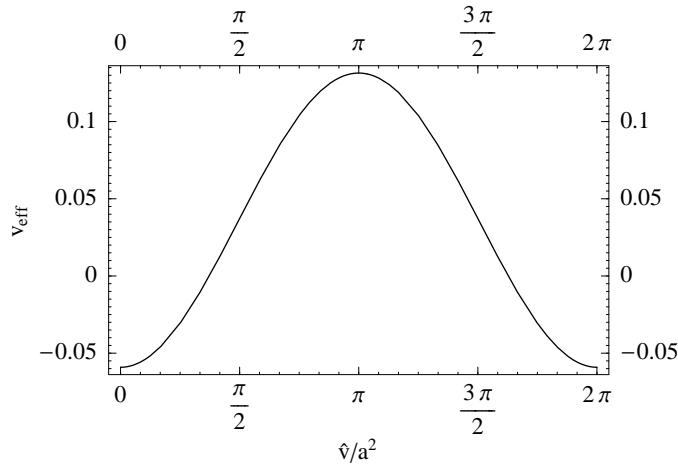,width=10cm}
    \caption{Normalized effective potential $v_\text{eff}$ vs $\vhat/a^{2}$
             with $a=10^{-15}$.}
    \label{fig_eff_pot}
    \end{center}
  \end{figure}

In Fig.~\ref{fig_eff_pot}, we plot $v_\text{eff}$ as a function of $\vhat/a^{2}$ when $a=10^{-15}$.
Contribution from the ghost loop is equal to Eq.~\eqref{eq_Veff_vec}
multiplied by $-2$.
The final result including gauge and ghost field contributions is therefore:
\begin{align}
  V_\text{eff} &= {3\over 32\pi^{2}}(ka)^{4}v_\text{eff}.
  \end{align}
  Note that if we include extra adjoint bulk fermions,
  they would contribute with opposite signs to Eq.~\eqref{eq_Veff_vec}
  and that if we add more than required to make the theory supersymmetric,
  the potential of Fig.~\ref{fig_eff_pot} would be flipped upside down,
  realizing a dynamical symmetry breaking vacuum
  which corresponds to the maximal twist $\theta=\pi/2$ in the $A_{\y}^{c}=0$ gauge.
This vacuum breaks $SU(2)$ completely and hence provides a rank reduction of the gauge symmetry.
We find that the symmetry breaking scale is of the order of $ka\simeq \text{TeV}$ for this case.

\section{Summary and discussions}
We have studied the $SU(N)$ pure gauge theory in the bulk of the Randall-Sundrum geometry
and have obtained Kaluza-Klein expansions of gauge and ghost fields 
under the presence of the gauge field background with most general twists $P_{i}$.
We find that four dimensional gauge, vectoscalar and ghost fields have exactly the same
KK masses.
During the course of this calculation we have clarified the notion of a large gauge transformation
that is non-periodic on the covering space and
how it is consistently realized in the warped background.
The effective potential for the background $A_{\y}^{c}$ is obtained.
We find that a gauge symmetry corresponding to a continuous Wilson line,
i.e.\ a $SU(2)$ subgroup of Eq.~\eqref{bc_mat},
which is completely broken for finite $\theta$ at the classical level,
is dynamically restored to $U(1)$.

It is straightforward to apply our method to include other fields
with or without extra boundary masses and especially to
supersymmetrize our setup, where the symmetry breaking scales due to
continuous Wilson lines will be of the order of $k a\simeq\text{TeV}$
according to the analysis presented here.
Since we find that the dynamics of Wilson lines in warped space can be controlled
in quite a parallel manner to that in flat space
if one treats the large gauge transformation carefully,
we expect that 
the Hosotani mechanism will lead to a dynamical supersymmetry breaking
of the order of weak scale in the setup discussed in the introduction.
It is also interesting to pursue the AdS/CFT correspondence
generalizing the analysis of Ref.~\cite{Contino:2003ve}
to our setup,
as the Wilson line on the AdS side
will correspond to a quantity that is integrated
all the way from UV to IR on the CFT side.
The techniques developed here can also be applied to the gauge-Higgs unification models in warped space.
These points will be presented in separate publications~\cite{KinyaAndi}.

\bigskip
\noindent{\bf Acknowledgments:}
We wish to acknowledge the helpful communication
with H.\ Abe, A.\ J.\ Buras, M.\ Drees, D.\ Ida, T.\ Kobayashi, A.\ Pomarol, 
A.\ Poschenrieder, M.\ Quiros, Y.\ Sakamura and A.\ Wingerter.
K.O.\ thanks N.\ Haba, Y.\ Hosotani and S.\ Komine for useful discussions.
The work of K.O.\ is partly supported by the SFB375 of the Deutsche
Forschungs\-gemein\-schaft and A.W. is partly supported by the German
Bundesministerium f\" ur Bildung und Forschung under the contract
05HT4WOA/3 and the DFG Project Bu.\ 706/1-2.

\renewcommand{\baselinestretch}{1}\normalsize

\providecommand{\href}[2]{#2}\begingroup\raggedright\endgroup


\end{document}